\documentclass[12pt]{article}
\usepackage{times}

\usepackage{amsmath,amssymb,amsfonts} 
\usepackage{subeqnarray}
\usepackage{empheq}
\usepackage{graphicx}                 
\usepackage{color}                    
\usepackage{hyperref}                 
\usepackage{makeidx}
\usepackage{makecell}    
\usepackage[backend=bibtex, style=authoryear, natbib=true, sorting=nyt, maxbibnames=99]{biblatex}
\usepackage{lineno}
\usepackage{changes}

\renewbibmacro{in:}{}

\usepackage{orcidlink}

\DeclareGraphicsExtensions{.jpg}

\makeindex

{\end{quote}}

\title{Fluid drag on a sphere at finite Reynolds number---a classical problem
examined afresh} 

\author
{James Q. Feng\\
\\
\normalsize{OXCBO Research, Maple Grove, Minnesota, USA}\\
\\
\normalsize{E-mail:  james.q.feng@gmail.com}
}

\date{}

\begin{document} 


\baselineskip24pt


\maketitle 



\begin{abstract}
\noindent Evaluating the drag coefficient $C_d$ for a sphere moving in a fluid 
at finite Reynolds number $Re$ has become 
a classical research subject in fluid dynamics, with publications appearing continuously 
by numerous authors over the years due to the unavailability of closed-form mathematical solutions. 
To date, mathematical formulas for calculating $C_d$ as a function of $Re$ can only be obtained 
by correlation curve-fitting to experimental data. 
The consequence is the coexistence of various formulas with different mathematical fitting functions
and an increasing number of fitting parameters.
The present study focuses on examining the accuracy of Klyachko's 1934 formula for $C_d(Re)$ 
with two common fractions as fitting parameters, which could enable 
closed-form integration of a simplified equation governing rectilinear motion of aerosol spheres. 
The computed values of root-mean-square relative error from 
recently developed formulas show reasonable accuracy of the Klyachko formula for $Re \le 1000$
and its extension to $Re < 2 \times 10^5$ as a piecewise continuous function,
with the possibility of significant improvement through a slight modification. 
Hence, a few demonstrative examples are presented using the extended Klyachko formula in the
study of rectilinear acceleration, the calculation of the terminal velocity of aerosol spheres, and so forth,
to shed fresh light on how this classical problem can be studied.
\end{abstract}


\section{Introduction}
As a fundamental canonical problem in fluid mechanics,  
flow past a sphere has been studied by numerous authors and discussed in 
well-known subject-matter books \citep{lamb1932, schlichting1955, landau1959, fuchs1964, clift1978, pruppacher1997}. 
Besides its theoretical convenience for the in-depth mathematical study of drag, boundary layer separation, 
and turbulence, the problem of a spherical body moving in a fluid often appears in a myriad of 
natural processes and 
engineering applications \citep{fuchs1964, clift1978, pruppacher1997}.
Despite the tremendous efforts by generations after generations of mathematically talented scientists, 
closed-form analytical solutions for flow around rigid and fluid spheres are effectively limited to 
very low values of the Reynolds number, $Re \equiv u d /\nu$  (where 
$d$ denotes the sphere's diameter, $u$ and $\nu$ the relative velocity and kinematic viscosity of the surrounding fluid). 
Knowledge about the situation at higher Reynolds numbers must resort to 
experimental measurements, numerical computations, as well as approximations based on boundary-layer theory  \citep{clift1978}.

Dictated by the laws of physics, a sphere (or any object of another shape) will experience 
a drag force---as the surface integral over its whole surface of all stresses arising from the relative fluid flow---whenever 
there is a relative motion of it in the surrounding fluid \citep{landau1959}.
In scientific literature, a dimensionless version of the drag force---the drag coefficient $C_d \equiv 8 F_d / (\pi d^2 \rho u^2)$ 
(with $F_d$ denoting the drag force and 
$\rho$ the density of surrounding fluid)---has commonly been
utilized for convenience of analytical description 
\citep{schlichting1955, landau1959, fuchs1964, clift1978, pruppacher1997, ramirez2017, goossens2019, michaelides2026}. 

The drag coefficient 
of a sphere in steady motion, in an incompressible Newtonian fluid, $C_d$
is often presented as a graph versus $Re$ known as the "Rayleigh curve" or ``standard drag curve'' \citep{clift1978}, 
which has been expressed in terms of mathematical functions with adjustable fitting parameters by various authors via
empirical or semiempirical means with non-stopping correlation efforts 
\citep[e.g.,][]{schiller1933, klyachko1934, lapple1940, brown2003, cheng2009},
including the usage of a genetic programming procedure \citep{barati2014} 
and machine learning with a symbolic regression algorithm \citep{hasadi2023}.
In view of the \citet{cheng2009} comparisons with compiled experimental data from multiple sources, 
the reported values of 
the ``average relative error''
of correlation formulas from different authors, with vastly different mathematical complexities, 
are mostly within $\sim 5\%$,
e.g., $3.019\%$ of \citet{clift1978}, $3.236\%$ of \citet{brown2003},  
$2.469\%$ of \citet{cheng2009}. and so on so forth.

Valuable as it might be for a percent or so improvement in the correlation relative error, 
publications by many recent authors 
\citep{brown2003, cheng2009, barati2014, kramer2021, hasadi2023} 
seem to focus only on  
sophisticated correlation formulas, often expressed in terms of transcendental functions 
with increasing number of the fitting parameters, for $0 < Re < 2 \times 10^5$, 
while completely omitting (or no more than barely mentioning) the much cleaner formula with only two fitting parameters, 
$C_d = 24 (1 + \alpha Re^\beta)/Re$,
suggested by \citet{schiller1933} (with $\alpha = 0.15$ and $\beta = 0.687$), 
\citet{klyachko1934} (with common fractions $\alpha = 1/6 \approx 0.167$ and $\beta = 2/3 \approx 0.667$), 
among others. 
According to \citet{fuchs1964}, the Klyachko formula appears quite accurate, 
agreeing with the experimental data 
at least for $3 < Re < 400$ (whereas \citet{klyachko1934} applied it for $0 < Re \le 1000$). 

Interestingly, a formula very similar to that of Klyachko (with only the parameter $\alpha =1/6$ replaced by $0.158$)
was adopted by \citet{serafini1954} (who claimed to agree with experimental values for 
$0.5 < Re < 500$ but without citing the previous work of Klyachko), 
in the derivation of a closed-form analytical solution.
Later, \citet{putnam1961} extended the validity of the Klyachko formula to $Re < 1000$ (also without citing 
prior publications by either Klyachko or Serafini), when demonstrating the existence of an analytical solution
(which had already been used by Klyachko in 1934). 
What has been implemented in the computational fluid dynamics (CFD) software package OpenFOAM$^{\circledR}$
for the ``sphere drag'' calculation is in terms of a piecewise continuous function
 \begin{equation}
 C_d (Re) =  \left\{  \begin{array}{rcl} 
  \frac{24}{Re}\left(1 + \frac{1}{6} Re^{2/3} \right) \quad , & \mbox{ for} & 0 < Re <Re_m = 1000 \\  \\
  \frac{24}{Re_m} \left(1 + \frac{1}{6} Re_m^{2/3} \right) = 0.424  \, , & \mbox{ for} & Re \ge Re_m = 1000 \,\, 
    \end{array}\right .  
  \,  ,
  \label{putnam}
\end{equation}      
following the KIVA-II work of \citet{amsden1989} (which would
actually be invalid beyond $Re \sim 2 \times 10^5$ when the abrupt ``drag crisis" occurs).
It is effectively an extension of the Klyachko formula for $Re$ up to $\sim 2 \times 10^5$ 
(
though Klyachko suggested $\sim 3 \times 10^5$ as the extension upper limit, which is realistically over-aggressive
and should be ignored
)  by continuously
connecting it to a constant of $4 (6 + Re_m^{2/3}) / Re_m$ at $Re = Re_m$, 
where the value of $Re_m$ is chosen as $1000$ by Amsden et al. in (\ref{putnam}). 

The mathematical function of (\ref{putnam}) for $0 < Re < 1000$, though significantly nonlinear, 
was shown to enable the derivation of a closed-form analytical solution to the differential equation
governing the rectilinear motion of a sphere in a fluid \citep{serafini1954, putnam1961}, 
when forces 
(such as added mass and Basset-Boussinesq history force)
other than the steady-flow drag are negligible, such as situations often encountered 
with aerosol particles \citep{fuchs1964}.  
A closed-form analytical solution in fluid dynamics of a moving sphere is 
generally considered intractable, and highly appreciated for its mathematical elegance 
even though it is useful only for special circumstances in a limited interval of $Re$ values.  
Therefore, it deserves special attention in the scientific community.  

In what follows, we start by examining the numerical differences between 
(\ref{putnam}) and those of the more recent formulas, 
to shed light on how good (or bad) the extended
Klyachko formula can be, thereby closing a knowledge gap in the scientific literature.
Then, a few examples are presented for practical applications using the $C_d(Re)$ given by (\ref{putnam})
or its improved version.

\section{Comparison of $C_d(Re)$ with a reference formula}
To measure the difference (or ``error'') between the extended Klyachko formula (\ref{putnam})
and a recently developed correlation formula, such as that of \citet{brown2003} or others, 
the root-mean-square relative error (RMSRE):
 \begin{equation}
 \mathrm{RMSRE} = \sqrt{\frac{1}{b - a} \int_a^b \left[\frac{f(x) - F(x)}{F(x)} \right]^2 dx}
  \,\,\,  ,
  \label{rmsre}
\end{equation}
is used here, where $f(x)$ denotes the test function and $F(x)$ the reference function.

 \begin{table}
  \begin{center} 
\caption{RMSRE values for $C_d(Re)$ of Klyachko's formula expressed as Eq. (\ref{putnam}) versus 
that obtained by \citet{brown2003}, 
\citet{cheng2009}, and \citet{barati2014} }
  \begin{tabular}{c|ccc}
  \hline\
    $Re$ interval &  Brown and Lawler & Cheng & Barati et al.   \\[3pt]
  \hline    
     ($0.01$, $1000$) & $0.0434$ & $0.0411$ & $0.0464$ \\
     ($0.01$, $900$) & $0.0373$ & $0.0332$ & $0.0396$ \\
     ($0.01$, $800$)  & $0.0315$ & $0.0256$ & $0.0337$ \\
     ($0.01$, $500$) & $0.0198$ & $0.0112$ & $ 0.0285$ \\
     ($1000$, $10^5$) & $0.0753$ & $0.0711$ & $ 0.0734$ \\
     ($1000$, $2 \times 10^5$) & $0.1109$ & $0.1206$ & $ 0.1155$ \\
     ($Re_m = 900$, $10^5$) & $0.0539$ & $0.0466$ & $ 0.0525$ \\
     ($Re_m = 900$, $2 \times 10^5$) & $0.0529$ & $0.0586$ & $ 0.0562$ \\
  \hline
  \end{tabular}
  \end{center}
\end{table}

Table 1 presents the values of RMSRE (\ref{rmsre}) with $f(x)$ for the $C_d(Re)$ in (\ref{putnam}) and
the reference function $F(x)$ for 
the correlation formulas for $Re < 2 \times 10^5$ by recent authors, i.e.,

{\small \[
C_d = \tfrac{24}{Re} (1 + 0.15 Re^{0.681}) + \tfrac{0.407 Re}{Re+8710} \qquad \mbox{by \citet{brown2003}} \, ,
\] }
{\small \[
C_d = \tfrac{24}{Re} (1 + 0.27 Re)^{0.43} + + 0.47 [1 - \exp(-0.04 Re^{0.38})] \quad \mbox{by \citet{cheng2009}} \, ,
\] }
{\small \[
C_d = 5.4856\times10^9\tanh(4.3774\times10^{-9}/Re)+0.0709\tanh(700.6575/Re)  \quad 
\] 
\[
+0.3894\tanh(74.1539/Re) - 0.1198\tanh(7429.0843/Re)   \quad 
\] 
\[
+1.7174\tanh[9.9851/(Re+2.3384)] + 0.4744  \quad  \mbox{by \citet{barati2014}} \, ,
\] }
integrated from 
$Re = 0.01$ to $Re = 1000$, $900$, $800$, and $500$ 
as well as $Re = 1000$ to $Re = 2 \times 10^5$.  Apparently, for $0.01 < Re < 1000$, the RMSRE value 
is consistently lowest for (\ref{putnam}) versus the correlation formula of \citet{cheng2009},
while the differences between (\ref{putnam}) and all three correlation formulas of \citet{brown2003, cheng2009, barati2014}
have RMSRE values less than $5\%$.  
Most of the differences appear to come from the interval of $800 < Re < 1000$,
because a close examination shows that (\ref{putnam}) exhibits a gradual deviation from the  ``standard curve'' of $C_d$ versus $Re$  
for $Re > 500$ up to by almost $8\%$ underestimation around $Re = 1000$ where $C_d \sim 0.46$ 
has been consistently measured in experiments.
But for $1000 < Re < 2 \times 10^5$
(where the experimental data vary from $C_d \sim 0.40$ in $2000 < Re < 10^4$ to $C_d \sim 0.47$ in 
$4 \times 10^4 < Re < 2 \times 10^5$), 
the RMSRE values for  
the simple approximation
$C_d = 0.424$ versus those of the sophisticated correlation formulas 
become slightly over $10\%$, coming mostly from the interval of $10^5 < Re < 2 \times 10^5$.
Nonetheless, these few percentage points of differences between (\ref{putnam}) and more recent correlation formulas 
appear quite insignificant when viewed in the graph of figure 1 (where the relatively low RMSRE case of Cheng is omitted
without loss of illustrative effectiveness).

It should be noted that table 1 also reveals the fact that a significant reduction of the RMSRE values 
can be achieved by simply moving the upper limit of the Klyachko formula 
in (\ref{putnam}) from $Re_m = 1000$ to $900$, for a modified formula
with improved accuracy (especially for $Re > 1000$ up to $Re \sim 2 \times 10^5$
with the constant value of $C_d$ for $Re > Re_m$ effectively increased from $0.424$ to $0.441$
),
 \begin{equation}
 C_d (Re) =  \left\{  \begin{array}{rcl} 
  \dfrac{24}{Re}\left(1 + \dfrac{1}{6} Re^{2/3} \right)  \quad \quad  ,    & \,  \mbox{ for} & 0 < Re < Re_m = 900 \\  \\
    {4} (6 +  Re_m^{2/3} ) / Re_m \approx 0.441  \, , & \mbox{ for} & Re_m = 900 \le Re <  2 \times 10^5  \,\, 
    \end{array}\right .  
  \,  .
  \label{putnamModified}
\end{equation}     

As a comparative reference, the RMSRE values calculated for Brown and Lawler versus Barati et al. 
and Cheng versus Barati et al. in the interval from $Re = 0.01$ to $1000$ are $\sim 0.0228$ and $0.0161$, respectively.
Obviously, a few percent of disagreements among even those recent correlation formulas appear inevitable.
Thus, the accuracy of either (\ref{putnam}) or (\ref{putnamModified}) can be quite acceptable.

\begin{figure}[ht] \label{fig1}
\includegraphics[clip=true,scale=0.70,viewport=50 90 590 420]{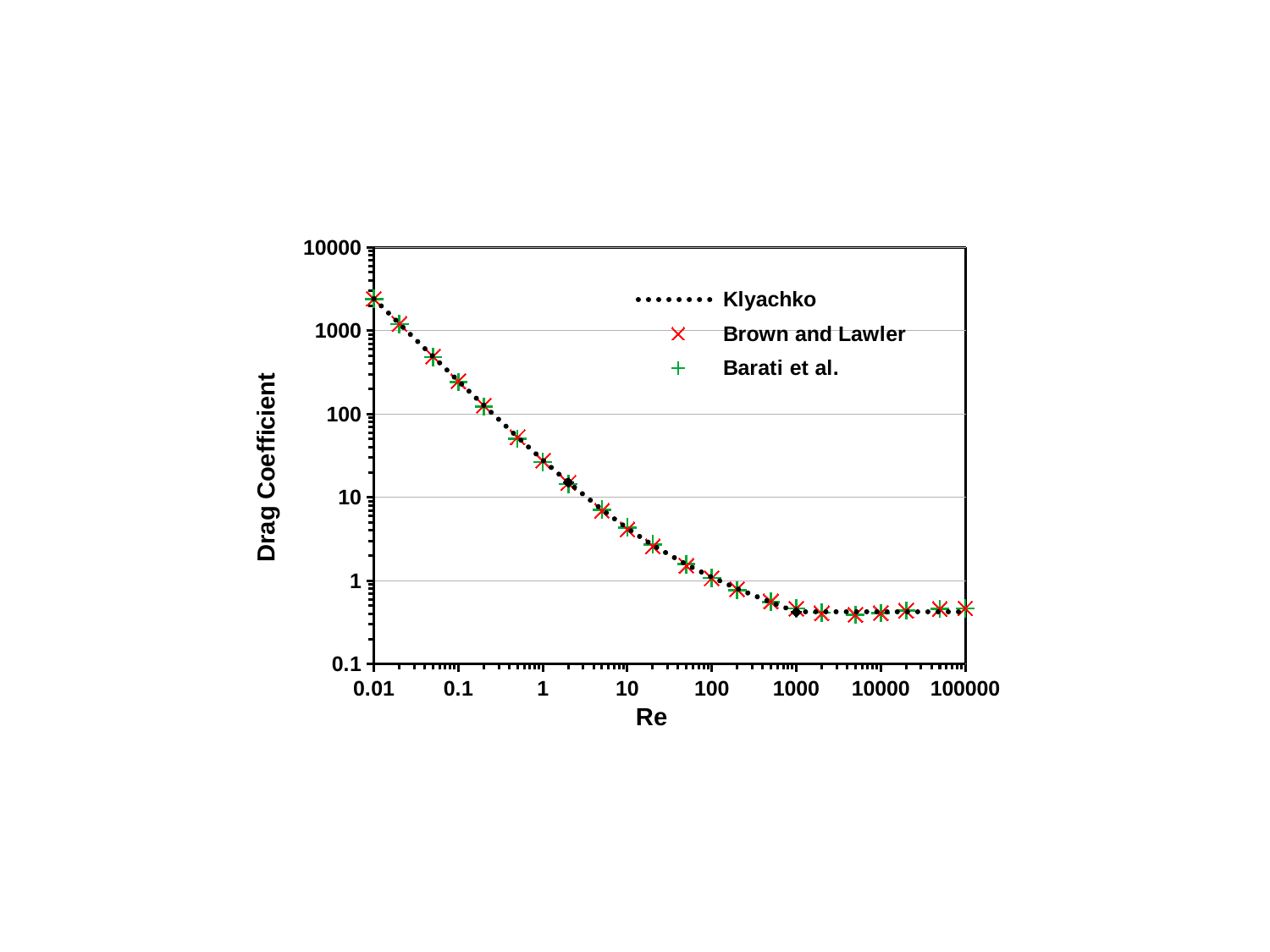}
\caption{Graphic comparison between Klyachko's formula expressed as Eq. (\ref{putnam}) and more recent correlation formulas for 
drag coefficient \citep{brown2003, barati2014} }
\end{figure}

In view of numerical differences among various available correlation formulas and the noisiness of experimental data,
the comparison between (\ref{putnam}) and recently developed ``more accurate'' correlation formulas 
appears quite reasonable for practical real-world applications, especially for $Re < 1000$. 
The ``accuracy'' of (\ref{putnam}) is shown here to further improve with a slight modification like (\ref{putnamModified}),
even without going through a process of rigorous optimization.

There seems to be no agreed-upon best metric for evaluating the correlation accuracy; different authors 
sometimes choose different sets of metrics in presentating correlation results.  
Even when RMSRE is used,  the formula (\ref{rmsre}) is but just one form for straightforward calculations, 
while it can have a corresponding discrete version as:
\[
\mathrm{RMSRE} = \sqrt{\frac{1}{N} \sum_{i=1}^N \left[\frac{f(x_i) - F(x_i)}{F(x_i)} \right]^2} \, ,
\]
the value of which could be sensitive to how those $N$ data points are distributed.  If so desired, those $N$ data points 
can be evenly distributed in a logarithmic scale of $Re$ as shown in figure 1 for discrete points of 
``Brown and Lawler'' and ``Barati et al.''  Then, the discrete RMSRE values of $Re$ in $(0.01, 1000)$ 
become $0.0290$, $0.0345$, $0.0375$ for (\ref{putnam}) versus ``Brown and Lawler'', ``Cheng'', ``Barati et al.'', respectively. 
For $Re$ in $(1000, 10^5)$, those discrete RMSRE values are $0.0685$, $0.0608$, $0.0687$. 
Despite slight numerical variations, 
whether to use (\ref{rmsre}) or its discrete version does not seem to make a
fundamental difference in the present case when comparing (\ref{putnam}) with other correlation formulas.

In terms of the discrete RMSRE, the value for the Brown and Lawler formula versus their set of $N = 480$ compiled experimental data is calculated as $0.0411$, and that for the Cheng formula as $0.0346$.  
On the other hand, \citet{brown2003, cheng2009, barati2014} also presented their correlation results using the same
set of experimental data in terms of the ``average relative error'', 
$\frac1N\sum (|C_{d, calc} - C_{d, exp}|/C_{d, exp})$, 
calculated as $0.0324$, $0.0247$, $0.0223$.  
No matter which metric is employed, the apparent accuracy improvements over the years seem to have approached their limit,
differing by no more than a few percent from inevitably noisy experimental data. 
Therefore, those recent correlation formulas can be reasonably regarded as trustworthy surrogates for 
the actual experimental data.   

\section{Particle rectilinear acceleration}  
For the case of quasistatic drag as the only dominant force controlling the rectilinear motion of a spherical particle 
(with particle density $\rho_p$ and velocity $u(t)$), 
Newton's law of motion based on (\ref{putnamModified})
for the ``initial particle Reynolds number'' $Re_0 \equiv u(0) d / \nu$ $< 2 \times 10^5$ can be written as
\begin{subequations}
  \label{u-ode}
\begin{empheq}[left={\dfrac{d {\tilde u}}{d {\tilde t}} =\empheqlbrace\,}]{align}
        &  - \tfrac{3 \rho (6 + Re_m^{\,\, 2/3})}{\rho_p \, Re_m} \, {\tilde u}^2 \approx \tfrac{0.331 \rho}{\rho_p} \, {\tilde u}^2 \, , \,\,  \mathrm{ for } \,\, 
900 \le Re_0 {\tilde u} < 2 \times10^5 \, , 
\label{u-ode0} \\ 
    &    - \frac{18 \, \rho}{Re_0 \, \rho_p}  \left(1 + \frac{Re_0^{\,\, 2/3}}{6} {{\tilde u}}^{\, 2/3} \right) {\tilde u}  \,\, ,
   \,\,  \mathrm{ for } \,\, 0 < Re_0 {\tilde u} < 900   \label{u-ode1}
\, \, ,  
\end{empheq}
\end{subequations} 
where  
${\tilde t} = t u(0) / d$ and ${\tilde u} = u / u(0)$.  

For the case of $Re_0 > Re_m = 900$, a straightforward integration of (\ref{u-ode0}) for ${\tilde u}(0) = 1$ 
yields
\begin{equation}
   {\tilde u}({\tilde t}) = \left[1 +
\left(\frac{0.331 \rho}{\rho_p} \right) {\tilde t} \right]^{-1} \, ,   
\,\,  \mathrm{ for } \,\,  {\tilde t} \le {\tilde t}_m = \frac{\rho_p}{0.331 \rho} \left(\frac{Re_0 - Re_m}{Re_m}\right)
\,  . 
  \label{u-int0}
\end{equation}
Then, for ${\tilde t} > {\tilde t}_m$, it turns out (though may not be immediately obvious) that  (\ref{u-ode1}) can be rearranged as
\[
   \frac{d \left({\tilde u}^{\,\, -2/3}\right)}{d {\tilde t}} = \frac{12 \, \rho}{Re_0 \, \rho_p}  \left({{\tilde u}}^{\,\, -2/3} + \frac{Re_0^{\, 2/3}}{6} \right) 
\quad ,
\]
which yields an analytical solution for ${\tilde u}({\tilde t}) < {\tilde u}({\tilde t}_m) = Re_m / Re_0$, similar to that shown
by \citet{serafini1954} and \citet{putnam1961} (equivalent to the case of $Re_m = Re_0$ 
such that ${\tilde t}_m = 0$ and $Re_m / Re_0 =1$), 
\begin{equation}
   {\tilde u}({\tilde t}) =  \left[
\left(\frac{Re_0^{\, 2/3}}{Re_m^{\, 2/3}} + \frac{Re_0^{\, 2/3}}{6} \right) \exp{\left(\frac{12 \, \rho}{Re_0 \, \rho_p} \Delta{\tilde t} \right)}  
      - \frac{Re_0^{\, 2/3}}{6} \right]^{ - 3/2}
     =\frac{ 6 \sqrt{6} }{Re_0 \, \Phi({\tilde t})^3}
\, \,  , 
  \label{u-int}
\end{equation}
where $\Delta{\tilde t} \equiv {\tilde t} - {\tilde t}_m$ and 
\[
\Phi ({\tilde t}) \equiv \sqrt{\left(\frac{6}{Re_m^{\, 2/3}} + 1 \right) 
 \exp{\left(\frac{12 \, \rho}{Re_0 \, \rho_p} \Delta{\tilde t} \right)}  - 1}
\quad .
\]
As \citet{serafini1954} pointed out, 
it is the exponent $2/3$ in (\ref{u-ode1}) that enables closed-form integration.
(In other words, the formula of \citet{schiller1933}, though very similar to that of Klyachko 
in mathematical form but with an exponent $\beta$ slightly deviating from $2/3$, 
cannot facilitate a closed-form integration in a rigorous sense.)
 
By integrating
 ${\tilde u}({\tilde t})$ given by (\ref{u-int0}) from $0$ to ${\tilde t}$ (if ${\tilde t} \le {\tilde t}_m$ for $Re_0 > Re_m$)
and (\ref{u-int}) from ${\tilde t}_m$ to ${\tilde t}$ (when ${\tilde t} > {\tilde t}_m$), 
the particle flight distance (in units of $d$) as a function of flight time can also be derived as
\begin{equation}
{\tilde z}({\tilde t}) =  \left\{  {\begin{array}{rcl}
        \dfrac{\rho_p}{0.331 \rho} \ln\left(1 + \dfrac{0.331 \rho}{\rho_p} {\tilde t}\right)  \quad , \quad \quad  \mathrm{ for } \,\, 
{\tilde t} \le {\tilde t}_m \, , \\ \\
  \dfrac{\rho_p}{0.331 \rho} \ln\left(1 + \dfrac{0.331 \rho}{\rho_p} {\tilde t}_m \right) 
 + \sqrt{6} \dfrac{\rho_p}{\rho } \biggl\{ \dfrac{1}{\Phi({\tilde t}_m)} - \dfrac{1}{\Phi(\Delta{\tilde t})} +  \\ \\
  + \tan^{-1} [\Phi ({\tilde t}_m)]  - \tan^{-1} [\Phi (\Delta{\tilde t})] \biggr\} \, , \,\,
 \mathrm{ for } \,\, 
{\tilde t} > {\tilde t}_m \, \, .
    \end{array}}\right .  
\end{equation}
Then, the stop distance (or the maximum achievable flight distance), 
often used to evaluate possible nozzle-to-substrate standoff in drop-on-demand inkjet printing for 
an ejected droplet,
when $Re_0 > Re_m$ becomes  
\begin{equation}
{\tilde z}(\infty) =  \dfrac{\rho_p}{0.331 \rho} \ln\left(1 + \dfrac{0.331 \rho}{\rho_p} {\tilde t}_m \right)
 + \sqrt{6} \frac{\rho_p}{\rho} \biggl\{ \frac{1}{\Phi({\tilde t}_m)} 
+ \tan^{-1} \left[ \Phi({\tilde t}_m)\right]  - \frac{\pi}{2} \biggr\}
   \, .
  \label{stop-distance}
\end{equation}
For the case of $Re_0 \le Re_m$, we should have $Re_m / Re_0$ replaced by $1$ 
and ${\tilde t}_m = 0$.  Thus (\ref{stop-distance})
can be written as \cite{serafini1954, putnam1961}
\begin{equation}
{\tilde z}(\infty) =  \sqrt{6} \frac{\rho_p}{\rho} \left[ \sqrt{\frac{Re_0^{\, 2/3}}{6}} 
+ \tan^{-1} \left(\sqrt{\frac{6}{Re_0^{\, 2/3}}}\right)  - \frac{\pi}{2} \right]  ,
\mathrm{ because } \, \,  \Phi(0) = \sqrt{\frac{6}{Re_0^{\, 2/3}} }
   \, .
  \label{stop-distance0}
\end{equation}

It should be noted that (\ref{u-ode}) and subsequent analytical formulas (\ref{u-int0})-(\ref{stop-distance0})
are valid only when $\rho_p >> \rho$ (as in the case of aerosols), 
such that the effects of time-dependent components of drag (
such as added mass and Basset-Boussinesq history force
included in the discussion of Landau and Lifshitz \cite{landau1959} as well as Fuchs \cite{fuchs1964})
are negligible,
and also when the effect of gravity could be ignored, 
i.e., the terminal velocity due to gravity is much smaller than the sphere velocity magnitude of interest,
as in the study of dust particles in dust-recollecting devices \citep{klyachko1934}, droplet impingement onto an airfoil 
\citep{serafini1954},
fuel droplet spray in a combustion chamber 
\citep{putnam1961, amsden1989}, among others.

\section{Particle terminal settling velocity}
With a given $C_d(Re)$ for a sphere moving in a fluid, 
the terminal velocity $u$ of a spherical particle with density $\rho_p$ settling under the influence of gravity 
can be readily calculated based on force balance:
 \begin{equation}
\frac{\pi d^2 \rho u^2 C_d}{8} = \frac{\pi d^3 (\rho_p - \rho) g}{6} \, \,
\mathrm{ or } \,\,
\frac{3 Re^2 C_d}{4} = d^3 \, \frac{g}{\nu^2}\left(\frac{\rho_p}{\rho}-1\right)
  \,\,\,  ,
  \label{f-balance}
\end{equation}
where $g$ denotes gravitational acceleration.  Noteworthy here is that
$3 Re^2 C_d/4$ is the Archimedes number (also called the Best number), 
while $Re^2 C_d$ is known as the Davies number, often used 
by atmospheric scientists in studying the terminal velocity of water droplets in air \citep{davies1945, beard1976}. 

If the particle diameter and velocity are nondimensionalized as
\begin{subeqnarray}
  \label{du-dless}
d_* = d \left[\frac{g}{\nu^2} \left(\frac{\rho_p}{\rho} - 1\right)\right]^{1/3} = \left(\frac{3 C_d Re^2}{4}\right)^{1/3}
  \quad  \mathrm{and} \slabel{du-dless0} \\
u_* = u \left[g \nu \left(\frac{\rho_p}{\rho} - 1\right)\right]^{- 1/3} = \left(\frac{4 Re}{3 C_d}\right)^{1/3}
  \quad  ,  \slabel{du-dless1}
\end{subeqnarray}
we have $u_* \times d_* = Re$ as suggested by \citet{brown2003}.

With $C_d$ given by (\ref{putnam}) or (\ref{putnamModified}), 
$d_*$ and $u_*$ can be easily calculated as functions of $Re$, as
 \begin{equation}
\begin{array}{lcl} 
 d_* (Re) =  \left\{  \begin{array}{rcl} 
  \left[{Re}\left(18 + 3 Re^{2/3} \right)\right]^{1/3} \, , & \mbox{ for} & 0 < Re < Re_m \\  \\
   \left(\frac{18 + 3 Re_m^{\, 2/3} }{Re_m}\right)^{1/3} Re^{2/3}  \quad , & \mbox{ for} & Re_m \le Re < 2 \times 10^5 \,\, 
    \end{array}\right .  
  \,  ,  \\ \\
\mbox{and then } \quad u_*(Re) = Re / d_*(Re)  \quad ,
\end{array}
  \label{d-Re}
\end{equation}      
where we have $Re_m = 1000$ with (\ref{putnam}) or $Re_m = 900$ with (\ref{putnamModified}), if so chosen.

However, most authors present curve-fitting formulas for directly calculting 
the terminal velocity from a given particle size, namely, $u_*(d_*)$ \citep{brown2003, cheng2009}.
In view of the fact that $Re = u_* \times d_*$ and $C_d$ as functions of $Re$, (\ref{du-dless0}) is equivalent to 
$d_* = f_d(u_* d_*)$ and (\ref{du-dless1}) $u_*=f_u(u_* d_*)$, none of which could readily be converted to 
an explicit function of $u_*(d_*)$.  
If (\ref{du-dless}) and (\ref{d-Re}) are used as a parametric representation 
for generating tabulated data or plot, complicated numerical inversion can be avoided. 
Occasionally though, explicit solutions of $u_*(d_*)$ can be obtained
when a very simplified form of $C_d(Re)$ was adopted, 
like that used by \citet{zigrang1981}, or the formula of (\ref{d-Re})
for $Re > Re_m$ which yields $u^* = \sqrt{Re_m \, d_* / (18 + 3 Re_m^{\,\, 2/3})}$, 
e.g., $u_* \approx 1.7733 \sqrt{d_*}$ for $Re_m = 1000$
 or $u_* \approx 1.7389 \sqrt{d_*}$ for $Re_m = 900$
where the difference seems negligible with either choice.

Obviously, calculating $d_*$ and $u_*$ from a given $Re$ according to  (\ref{d-Re})
 can be much more straightforward and less cumbersome; 
only a mindset is needed here for the calculation procedure starting with a given value of $Re$. 
In generating a lookup table or graph of $u_*$ versus $d_*$ for practical usage, 
which calculation procedure to follow usually does not make any difference. 

 \begin{table}
  \begin{center} 
\caption{Parametrically calculated values of $d_*$ and $u_*$ based on Eq. (\ref{d-Re}) 
for $Re_m = 1000$ from given $Re$ compared with 
the correlation formulas of $u_*(d_*)$ obtained by \citet{brown2003}, 
and \citet{cheng2009} }
  \begin{tabular}{c|cccc}
  \hline\
    $Re$ & $d_*$ & $u_*$ & Brown and Lawler & Cheng   \\[3pt]
  \hline    \
     $0.01$ & $0.56607$ & $0.017666$ & $0.0190347$ & $0.0177829$ \\
     $0.03$  & $0.818670$ & $0.036645$ & $0.0392648$ & $0.0371126$ \\
     $0.1$ & $1.23083$ & $0.081246$ & $0.0856651$ & $ 0.0832422$ \\
     $0.3$ & $1.79705$ & $0.166941$ & $0.172253$ & $ 0.173591$ \\
     $1$ & $2.75892$ & $0.362460$ & $0.365645$ & $ 0.380909$ \\
     $3$ & $4.17400$ & $0.718734$ & $0.720300$ & $ 0.743474$ \\
     $10$ & $6.83454$ & $1.463156$ & $1.494056$ & $1.454615$ \\
     $30$ & $11.21065$ & $2.676026$ & $2.813076$ & $ 2.629949$ \\
     $100$ & $20.21706$ & $4.946319$ & $5.226117$ & $ 5.001796$ \\
     $300$ & $35.76093$ & $8.389044$ & $8.438702$ & $ 8.609237$ \\
     $1000$ & $68.25624$ & $14.65067$ & $13.15452$ & $14.07184$ \\
     $3000$ & $141.9782$ & $21.13000$ & $20.23183$ & $ 21.4360$ \\
    $1 \times 10^4$ & $316.8163$ & $31.56403$ & $31.10793$ & $ 31.49032$ \\
    $3 \times 10^4$ & $659.0044$ & $45.52321$ & $45.33656$ & $ 44.28541$ \\
     $1 \times 10^5$ & $1470.531$ & $68.00265$ & $68.03615$ & $ 65.02726$ \\
  \hline
  \end{tabular}
  \end{center}
\end{table} 
 
Table 2 and figure 2 exemplify tabulated values of $d_*$, $u_*$ generated based on (\ref{d-Re}) for $Re_m = 1000$
in comparison with the correlation formulas:

{\small \[
u_* = \left[\left(\frac{18}{d_*^2}\right)^{0.898 (0.936 d_* + 1)/(d_* + 1)} + \left(\frac{0.317}{d_*}\right)^{0.449}\right]^{-1.114}  \qquad \mbox{by \citet{brown2003}} \, ,
\] }
{\small \[
u_* = \sqrt{\frac{4 d_*^4}{1296 (1 + 0.022 d_*^3)^{0.54} + + 1.41 d_*^3 [1 - \exp(-0.15 d_*^{\, 0.45})]}} \qquad \quad \mbox{by \citet{cheng2009}} \, ,
\] }
and the corresponding graph of $u_*$ versus $d_*$. 
Again not surpringly, the results of $d_*(Re)$ given by (\ref{d-Re}) for $Re_m = 1000$, based on 
as a slightly less accurate $C_d(Re)$ formula (\ref{putnam}),
are still in reasonable agreement with those of more recent correlation formulas.

\begin{figure}[ht] \label{fig2}
\includegraphics[clip=true,scale=0.68,viewport=50 90 620 420]{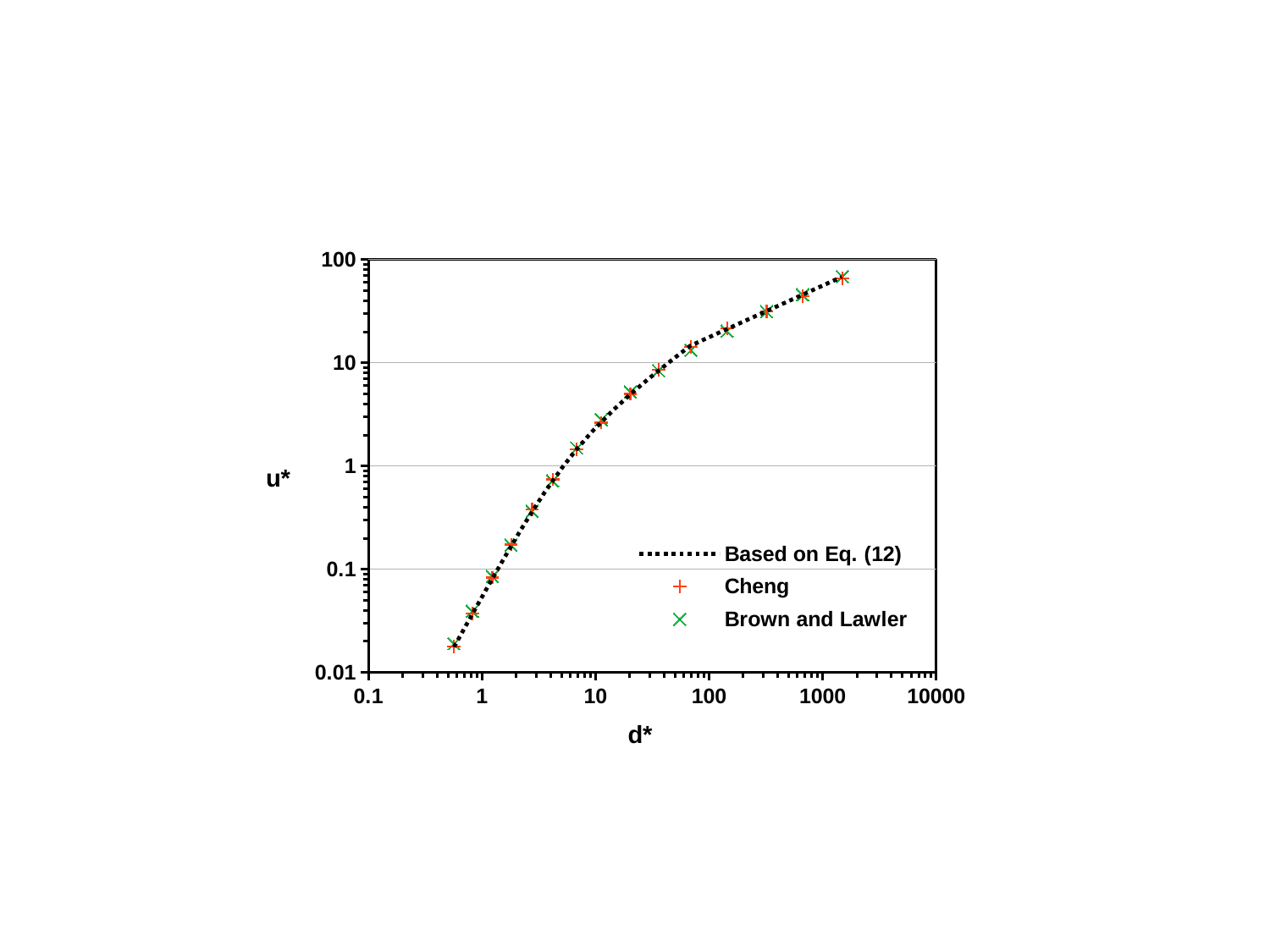}
\caption{Graphic comparison between parametrically calculated $u_*$ versus $d_*$ based on Eq. (\ref{d-Re}) for $Re_m = 1000$
and more recent correlation formulas of 
$u_*(d_*)$ by \citet{brown2003, cheng2009} }
\end{figure}

One of the most studied cases of terminal velocity is a water droplet falling in air, 
with $\rho_p/\rho = 1000/1.2$, $\nu = 1.5 \times 10^{-5}$ (m$^2$ s$^{-1}$), and $g = 9.81$ (m s$^{-2}$),
because of its direct relevance to cloud physics and
rainfall rate in meteorology. 
Thus, we have $d = 0.030205 \times d_*$ (mm) and $u = 0.49661 \times u_*$ (m/s), 
according to (\ref{du-dless0}) and (\ref{du-dless1}). 
For example, at $Re = 270$ (or $70$), eq. (\ref{d-Re}) yields $d^* \sim 33.82$ (or $16.90$) corresponding to 
$d \sim 1.02$ (or $0.51$) mm,  and $u^* \sim 270/33.82 = 7.98$ (or $70/16.90 = 4.14$) 
with $u \sim 3.96$ (or $2.06$) m/s, fairly close to the measured values 
tabulated by \citet{gunn1949},  
e.g., for $d = 1$ (or $0.5$) mm, $u = 403$ (or $2.06$) m/s with calculated $Re = 269$ (or $68.7$).

Another important problem closely related to terminal velocity is the dynamical process of the acceleration of a sphere
toward its terminal velocity, often needed for evaluating the required free fall length in an experimental setup \citep{wang1977}, 
among other applications. 
Similar to that with (\ref{u-ode}), assuming
quasistatic drag but with the addition of a term representing the gravitational force 
in the equation for rectilinear motion  
(for $Re < Re_m $):
\begin{equation}
   \frac{d {\tilde u}}{d {\tilde t}} = \frac{3 \rho \, (6 + Re^{2/3})}{\rho_p \, Re}  \left(1 - \frac{6 {\tilde u} + Re^{\, 2/3} {\tilde u}^{5/3}}
{6 + Re^{\, 2/3}} \right)
\, \, , \mathrm{ with  } \, \, Re \equiv \frac{u(\infty) \, d}{\nu} 
   \quad  
  \label{u_t-ode}
\end{equation}
denoting the particle Reynolds number for terminal velocity $u(\infty)$, where $d$ is the sphere diameter, 
with $0 \le {\tilde t} = t u(\infty) / d < \infty$ and $0 \le {\tilde u} = u / u(\infty) \le 1$.  
Here, 
\begin{equation}
 \frac{3 \, \rho (6 + Re^{2/3})}{\rho_p \, Re} = \frac{\rho g d}{\rho_p u(\infty)^2} \left(\frac{\rho_p}{\rho}-1\right)
  \label{u_t}
\end{equation}
reflects the fact of gravitational-force-driven motion.

Although differing from (\ref{u-ode}) only by a constant term,  (\ref{u_t-ode}) does not admit 
a closed-form analytical solution.
However, it would not be difficult to compute 
numerical solutions of ${\tilde u}({\tilde t})$ by 4th-order Runge-Kutta integration of (\ref{u_t-ode}),
and then ${\tilde z}({\tilde t})$ by integration of ${\tilde u}$, from $0$ to ${\tilde t}$, 
for generating results for ${\tilde u}$ (in units of $u(\infty)$) versus ${\tilde z}$ (in units of $d$).  

In the case of $Re > Re_m$, (\ref{u_t-ode}) needs to be rewritten as
\begin{subequations}
  \label{u_t-ode1}
\begin{empheq}[left={\dfrac{d {\tilde u}}{d {\tilde t}} =\dfrac{3 \rho \, (6 + Re_m^{2/3})}{\rho_p \, Re_m} \times \empheqlbrace\,}]{align}
        &  1 - \left(\frac{Re_m}{Re} \right)  \left(\frac{6 {\tilde u} + Re^{\, 2/3} {\tilde u}^{5/3}}{6 + Re_m^{\, 2/3}} \right) \, , \,  
\mathrm{ for } \, \,  0 < {\tilde t} \le {\tilde t}_m  \, , 
\label{u_t-ode10} \\ 
    &  \qquad \qquad  1 - {\tilde u}^2 \qquad  \qquad \qquad , \,\, 
 \mathrm{ for } \, \,  {\tilde t} \ge {\tilde t}_m  \, ,
\label{u_t-ode11} 
\end{empheq}
\end{subequations} 
where (\ref{u_t-ode11}) for ${\tilde t} > {\tilde t}_m$ leads to an analytical solution
\begin{equation}
  {\tilde u}({\tilde t}) = \tanh\left[\frac{3 \, \rho (6 + Re_m^{\,\, 2/3})}{\rho_p \, Re_m} ({\tilde t} - {\tilde t}_m) 
+ \tanh^{-1} \left(\frac{Re_m}{Re}\right) \right] 
\, .
  \label{u_t11}
\end{equation}
 Here, the value of ${\tilde t}_m$ is determined by numerically integrating (\ref{u_t-ode10}) 
with respect to ${\tilde t}$ until ${\tilde u} = Re_m/Re$.
Thus, at ${\tilde t} = {\tilde t}_m$ we have 
$d{\tilde u} / d{\tilde t} = 3 \rho (6 + Re_m^{2/3})(Re^2 - Re_m^2) / (\rho_p Re_m Re^2)$.

Integrating the ${\tilde u}({\tilde t})$ of (\ref{u_t11}) from ${\tilde t}_m$ to ${\tilde t}$ yields the particle flight distance
\begin{equation}
  {\tilde z}({\tilde t}) =  {\tilde z}({\tilde t}_m) + \frac{\rho_p \, Re_m}{3 \, \rho (6 + Re_m^{\,\, 2/3})} 
\ln \left\{\frac{\cosh\left[\tfrac{3 \, \rho (6 + Re_m^{\,\, 2/3})}{\rho_p \, Re_m} ({\tilde t} - {\tilde t}_m) 
+ \tanh^{-1} \left(\tfrac{Re_m}{Re}\right) \right]}{\cosh\left[\tanh^{-1}(Re_m/Re)\right]}  \right\}
\, .
  \label{z_t11}
\end{equation} 

Because $d {\tilde u} / d{\tilde t}$ $= {\tilde u} d{\tilde u}/d{\tilde z}$, an analytical formula explicitly  
relating ${\tilde z}$ to ${\tilde u}$ (or ${\tilde u}$ to ${\tilde z}$) can also be obtained, for ${\tilde t} > {\tilde t}_m$ when
${\tilde u} > Re_m / Re$,  based on (\ref{u_t-ode11}), e.g.,
\begin{equation}
  {\tilde z}({\tilde u}) =  {\tilde z}(Re_m/Re) + \frac{\rho_p \, Re_m}{6 \, \rho (6 + Re_m^{\,\, 2/3})} 
\ln \left[\frac{1 - (Re_m / Re)^2}{1 - {\tilde u}^2} \right]
\, ,
  \label{z_u11}
\end{equation} 
where the value of ${\tilde z}(Re_m/Re)$ should be determined by numerical integration of (\ref{u_t-ode10}) 
from ${\tilde t} = 0$ to ${\tilde t}_m$ though.

\begin{figure}[ht] \label{fig3} 
\includegraphics[clip=true,scale=0.66,viewport=40 80 650 450]{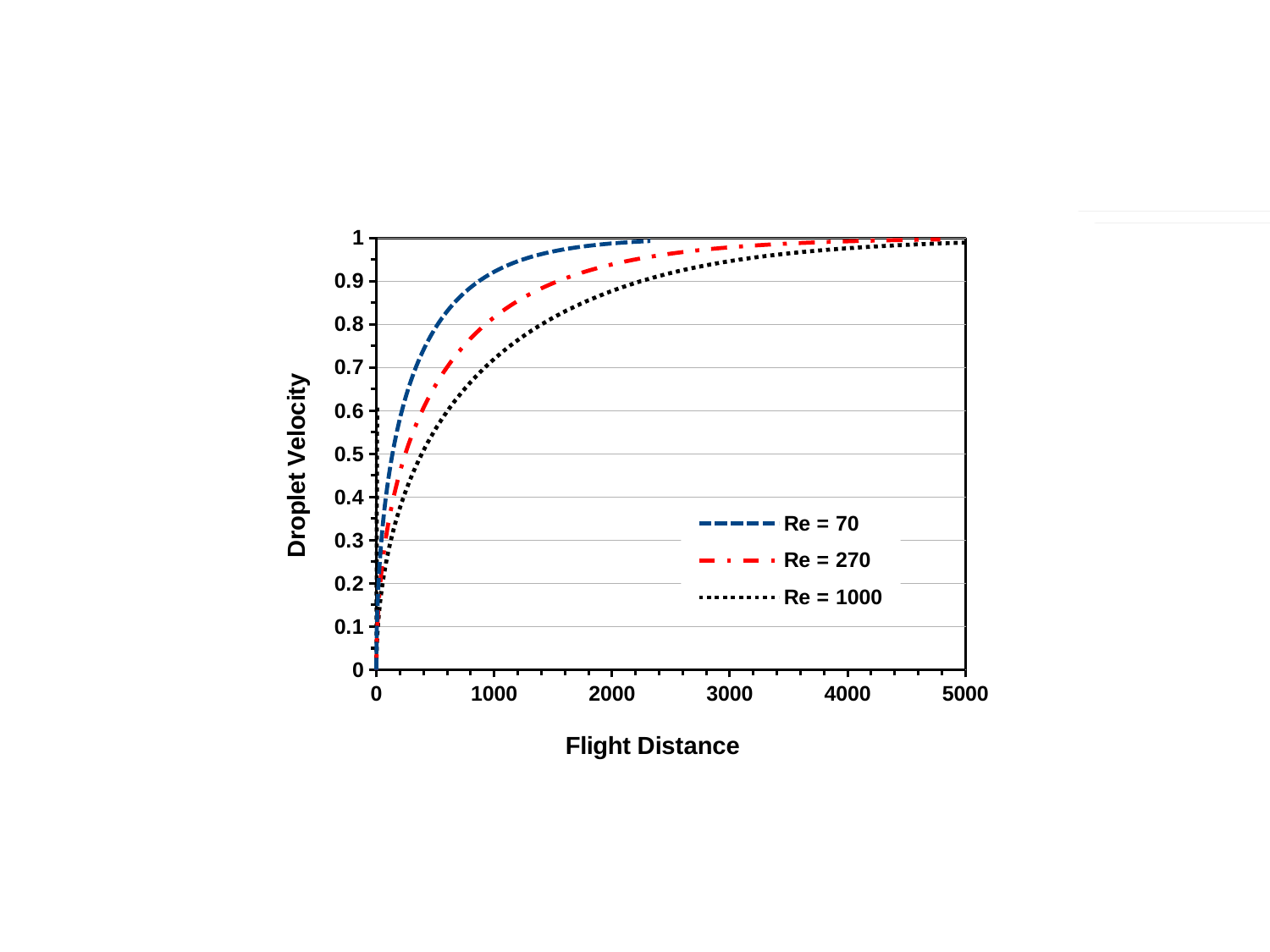}
\caption{Graphic display of ${\tilde u}({\tilde t})$ versus ${\tilde z}({\tilde t})$ generated by numerically integrating (\ref{u_t-ode}), 
for $Re = 70$ and $270$ (when $Re < Re_m$), and 
(\ref{u_t-ode1}), for $Re = 1000 > Re_m = 900$,
and then integrating ${\tilde u}({\tilde t})$ for ${\tilde z}({\tilde t})$ } 
\end{figure}
 
Figure 3 provides but one graphic example of ${\tilde u}$ versus ${\tilde z}$ for $Re = 70$, 
$270$, and $1000$ (for $Re_m = 900$), corresponding to $d \sim 0.51$, $1.02$, and $2.09$ mm of spherical particles 
of density $\rho_p \sim 1000$ kg m$^{-3}$ falling in air with terminal velocity $u(\infty)$
$\sim 2.06$, $3.96$, and $7.18$ m/s, agreeing well with experimental data for water droplets 
falling in air \citep{gunn1949, beard1976}, especially for $d < 1$ mm when droplet deformations 
are negligible. 

In a strictly theoretical sense, it takes infinite time and length for a particle to reach its terminal velocity.
But evaluating a finite falling length for ${\tilde u}$ to reach $0.99$ or $0.95$, 
sometimes denoted as ${\tilde z}_{99}$ or  ${\tilde z}_{95}$, could be more useful for 
practical applications \citep{wang1977}.
Thus, in the present context (for $Re = 70$, $270$, and $1000$), we have
${\tilde z}_{99} \sim 2139.58$, $3765.22$, and $5092.97$, corresponding to
$z_{99} \sim 1.09$, $3.85$, and $10.64$ m (for spherical particles of $d \sim 0.51$, $1.02$, and $2.09$ mm
of density $\rho_p \sim 1000$ kg m$^{-3}$).  
As expected, the computed values of $z_{99}$ here for $d < 1$ mm are in good agreement with those presented by 
\citet{wang1977}
shown with a curve plotted in their Fig. 3 for $d \sim 0.5$ and $1$ mm, $z_{99} \sim 1$ and $4$ m.
If ${\tilde u}({\tilde z}_{95}) = 0.95$ is practically acceptable, we could have considerably reduced 
fall distances for measurement setups, as $z_{95} \sim 0.635$, $2.24$, and $6.46$ m 
(corresponding to ${\tilde z}_{95} \sim 1243.43$, $2195.26$, and $3092.04$).
Apparently, we have $z_{95} / z_{99} < 0.61$ in this case.

\section{Summary}
Out of scientific curiosity about the less-known 1934 Klyachko formula for the sphere drag coefficient,
which exhibits an exceptionally clean mathematical expression,
an overdue relative-error evaluation is performed by comparing it with more recent correlation formulas,
for quantitatively understanding how good (or bad) this early curve-fitting formula,
as well as its extension to the Reynolds number $Re < 2 \times 10^5$ as a piecewise continuous function,
can be for practical applications. 
The results indicate quantitatively rather insignificant differences between the extended Klyachko formula $C_d(Re)$ 
and other apparently more sophisticated correlation formulas, for $Re < 1000$ (or, 
better yet, for $Re < 900$). 

Once the confidence is established for its reasonable accuracy, the extended Klyachko formula $C_d(Re)$
is briefly shown, similar to \citet{serafini1954} and \citet{putnam1961},
to yield closed-form analytical solutions (covering the applicable range of $0 < Re < 2 \times 10^5$) 
for the study of rectilinear acceleration of aerosol spheres, 
when the effect of gravity can be ignored. 
Then, a demonstration of using the extended Klyachko formula $C_d(Re)$ is presented for conveniently calculating
the spherical particle diameter and terminal settling velocity as functions of $Re$,  
yielding results in good agreement with recent correlation formulas of 
\citet{brown2003} and \citet{cheng2009}.
As expected, the dynamical process of an aerosol sphere accelerating toward its terminal velocity 
can also be described by solutions of ordinary differential equations based on 
the extended Klyachko formula, 
with comparable results to such relevant data of \citet{wang1977} 
for water droplet falling in air when 
droplet deformations are negligible.
These demonstrative examples are expected to enhance confidence in
the extended Klyachko formula for conveniently studying the classical problem
of aerosol (sphere) dynamics, among other applications.  
The derived closed-form analytical formulas can also be applied in modern CFD code development and 
validation.

\section*{Acknowledgments}
The author would like to thank Prof. Rodney Kubesh  (of St. Cloud State University) for helpful discussions and Prof. Carlos Ramírez (of University of Puerto Rico) for 
providing a copy of the original 1934 Russian article by Klyachko.

\section*{Funding}
None

\section*{Ethics declarations}
\subsection*{Conflict of interests}
The author has no conflicts to declare

\section*{ORCID iD}
James Q Feng  \orcidlink{0000-0003-4041-3179} 0000-0003-4041-3179

\printbibliography

\end{document}